# Blocked radiative heat transport in the hot pyrolitic lower mantle

Sergey S. Lobanov[1,2,3,*], Nicholas Holtgrewe[1,4], Gen Ito[2], James Badro[5], Helene Piet[6], Farhang Nabiei[6,7], Jung-Fu Lin[8], Lkhamsuren Bayarjargal[9], Richard Wirth[3], Anja Schreiber[3], Alexander F. Goncharov[1,10]

[1]Geophysical Laboratory, Carnegie Institution of Washington, Washington, DC 20015, USA

[2]Department of Geosciences, Stony Brook University, Stony Brook, NY 11794, USA

[3]Helmholtz Center Potsdam, GFZ German Research Center for Geosciences, Telegrafenberg, 14473 Potsdam, Germany

[4]Center for Advanced Radiation Sources, University of Chicago, IL 60637, USA

[5]Institut de Physique du Globe de Paris, Sorbonne Paris Cite, Paris 75005, France

[6]Earth and Planetary Science Laboratory, Ecole Polytechnique Fédérale de Lausanne, CH-1015 Lausanne, Switzerland

[7]Centre Interdisciplinaire de Microscopie Electronique, Ecole Polytechnique Fédérale de Lausanne, CH-1015 Lausanne, Switzerland

[8]Department of Geological Sciences, Jackson School of Geosciences, The University of Texas at Austin, Austin, TX 78712, USA

[9]Institute of Geosciences, Goethe University Frankfurt am Main, Frankfurt am Main, Germany

[10]Key Laboratory of Materials Physics, Institute of Solid State Physics CAS, Hefei 230031, China

[*]E-mail: slobanov@gfz-potsdam.de

**Abstract**

The heat flux across the core-mantle boundary ($Q_{CMB}$) is the key parameter to understand the Earth's thermal history and evolution. Mineralogical constraints of the $Q_{CMB}$ require deciphering contributions of the lattice and radiative components to the thermal conductivity at high pressure and temperature in lower mantle phases with depth-dependent composition. Here we determine the radiative conductivity ($k_{rad}$) of a realistic lower mantle (pyrolite)[1] *in situ* using an ultra-bright light probe and fast time-resolved spectroscopic techniques in laser-heated diamond anvil cells. We find that the mantle opacity increases critically upon heating to ~3000 K at 40-135 GPa, resulting in an unexpectedly low radiative conductivity decreasing with depth from ~0.8 W/m/K at 1000 km to ~0.35 W/m/K at the CMB, the latter being ~30 times smaller than the estimated lattice thermal conductivity at such conditions[2,3]. Thus, radiative heat transport is blocked due to an increased optical absorption in the hot lower mantle resulting in a moderate CMB heat flow of ~8.5 TW, at odds with present estimates based on the mantle and core dynamics[4,5]. This moderate rate of core cooling implies an inner core age of about 1 Gy and is compatible with both thermally- and compositionally-driven ancient geodynamo.

**Main text**

Heat exchange rate between the mantle and core ($Q_{CMB}$) is of primary importance for mantle convection and core geodynamo, the two processes that have been paramount for life on Earth. Considerations of the mantle and core energy budgets suggest a $Q_{CMB}$ in the range of 10-16 TW (*e.g.* Ref.[5]). Independently, transport properties of the mantle can provide insights into the $Q_{CMB}$ as it is controlled by the thermal conductivity of the mantle rock at the core-mantle boundary (CMB). Total thermal conductivity of primary lower mantle minerals is a sum of its lattice and radiative components ($k_{total} = k_{lat} + k_{rad}$). At near-ambient temperatures ($T$ ~300 K) the dominant mechanism of heat conduction is lattice vibrations while radiative transport is minor. At high temperature, however, the radiative mechanism is expected to become much more effective[6] as $k_{rad}(P,T) \sim \frac{T^3}{\alpha(P,T)}$ (Eq.1), where $\alpha(P,T)$ is the pressure- and temperature-dependent light absorption coefficient of the conducting medium. Mantle $k_{rad}$ is expected to increase with depth and light radiation might even be the dominant mechanism of heat transport in the hot thermal boundary layer (TBL) a few hundred km above the core[7,8]. To reconstruct mantle radiative thermal conductivity one needs to know the absorption coefficient of representative minerals in the near-infrared (IR) and visible (VIS) range collected at *P-T* along the geotherm.

Light diffusion in the hot lower mantle is governed by absorption mechanisms in iron-bearing bridgmanite (Bgm), post-preovskite (Ppv), and ferropericlase (Fp) as these minerals have absorption bands in the near-infrared (IR) and visible (VIS) range[7,9-11]. The intensity and position of absorption bands in the spectra of iron-bearing minerals vary with pressure[7,9-11] and temperature[12-14]. For example, previous room-temperature studies identified an increase in the absorption coefficient of Fp up to 135 GPa associated primarily with the spin transition and red-shift of the Fe-O charge transfer band[15,16]. Likewise, studies of upper mantle minerals at moderate temperatures have recognized substantial variations in their absorption spectra in the IR-VIS range. Crystal field and Fe-O charge transfer bands in olivine show an apparent intensification and broadening upon heating to 1700 K at 1 atm[12]. Similarly, characteristics of the $Fe^{2+}$-$Fe^{3+}$ charge transfer bands in wadsleyite and ringwoodite are altered upon heating to ~800 K (Ref.[13]). Although these transformations cannot be reliably extrapolated to temperatures of several thousand Kelvin, similar absorption mechanisms govern the optical properties of Bgm, Fp, and Ppv in the lower mantle, and thus must affect their radiative conductivity. Despite the importance, optical properties of Bgm, Ppv, and Fp have never been measured at mantle $P$-$T$ conditions as such measurements are challenging because of the thermal radiation interfering with the probe light at high $T$. On top of the $P$- and $T$- dependence of optical properties, absorption coefficients are sensitive to iron concentration, its valence/spin state, and crystallographic environment. In the mantle all of these vary with depth in a complex and not yet fully understood way[1], contributing to the uncertainty in the mantle $k_{rad}$. As a result of these complications, current $k_{rad}$ models are exclusively based on room-temperature measurements of Bgm and Fp with fixed chemical composition[7,9-11].

In this work, we performed optical measurements at high $P$-$T$ in laser-heated diamond anvil cells (DACs) using an ultra-bright light probe synchronized with fast time-resolved IR and VIS detectors that allowed us to diminish the contribution of thermal background (Methods). We determine the radiative conductivity of the lower mantle by measuring the spectral optical depth in pyrolite, a chemical proxy of the mantle[1], along the Earth's geotherm. Samples of highly homogeneous pyrolite glass (Extended Data Table 1) were crystallized at the thermodynamic conditions of the lower mantle producing a conglomerate of Bgm (±Ppv) and Fp with compositions representative of the equilibrium in the lower mantle (Fig. 1 and Methods). The latter allowed us to account for the effects of $P$- and $T$-dependent crystal chemistry such as the iron content and its valence/spin states in the lower mantle phases.

Prior to optical measurements, each glass sample was crystallized at $P > 30$ GPa and $T$ up to 3000 K for several minutes. Full glass crystallization was confirmed by the disappearance of a diffuse low frequency Raman peak (boson peak), which is characteristic of glasses (Extended Data Fig. 1). The mineralogical content of all crystalized samples is dominated by Bgm (+ Ppv at

~135 GPa) and Fp as revealed by the synchrotron x-ray diffraction (XRD) (Extended Data Fig. 2). In addition to XRD characterization, one of the recovered pyrolite samples (~2800 K and 56 GPa) was further analyzed using scanning transmission electron microscopy (STEM) and energy-dispersive x-ray spectroscopy (EDX) in order reveal its fine texture and the chemical composition of constituting minerals (see details on analytical techniques in Methods). High-angle annular dark-field (HAADF) images show that the average grain size of the crystallized pyrolite is smaller than ~500 nm (Fig. 1), suggesting that light scattering on grain boundaries might be substantial[17] and needs to be taken into account. The bulk of the crystallized sample is composed of Bgm, Fp, and Ca-perovskite in the proportions that are consistent with the previous reports on pyrolite mineralogy (~75, 18, and 7 vol.%, respectively[1]). Coexisting Bgm and Fp are homogeneous in their chemical composition and show no significant variations across the probed areas with $Bgm^{Fe\#} = 7.7 \pm \sim2$ and $Fp^{Fe\#} = 16.8 \pm \sim2$ (Extended Data Table 2), also in agreement with that reported in the literature (*e.g.* Ref.[1]). In addition to Bgm, Fp, and Ca-perovskite, we have also observed scarce grains of metallic Fe (< ~1 vol.%), supporting the notion that the lower mantle may contain a small amount of metallic iron-rich alloy.

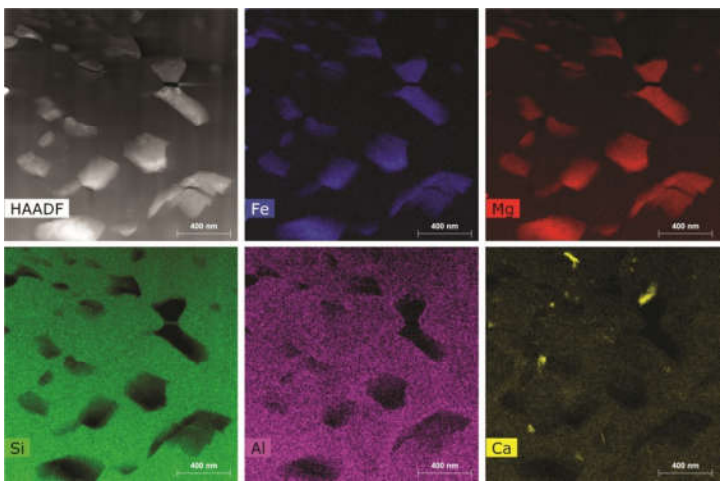

**Figure 1.** STEM HAADF image and elemental EDX maps of a pyrolite sample laser-heated at 56 GPa up to ~ 2800 K (see also Extended Data Fig. 3 for the optical image of this sample). Ferropericlase grains (Si, Al-free) are embedded in the Fe, Al-bearing bridgmanite matrix (Si-rich areas). Ca-perovskite is also present. The scale bar is 400 nm.

Upon heating, our IR-VIS measurements reveal that the optical absorbance of pyrolite shows a continuous strong increase up to the maximum temperature of ~2855 K at all studied pressures (Extended Data Fig. 4 and 5). Importantly, the observed temperature-enhanced absorption is reversible, as is indicated by the similarity of absorption coefficients measured prior and after the laser-heating cycles, suggesting that the increased opacity at high *T* is due to a temperature-activated absorption mechanism in the IR-VIS range. Figure 2 shows pyrolite absorption coefficient at selected *P-T* conditions that approximate the Earth's geotherm with the exception of the spectrum measured at 134 GPa and 2780 K. The absorption coefficient of pyrolite increases by a factor of 4-8 (depending on the frequency) from 40 GPa / 2120 K to 134 GPa / 2780 K. An even larger absorption coefficient might be expected for the CMB temperatures of ~4000 K. Accordingly, we extrapolated the 134 GPa absorption coefficient to 4000 K using its near-linear temperature dependence established in the 1700-2700 K range (Extended Data Fig. 6).

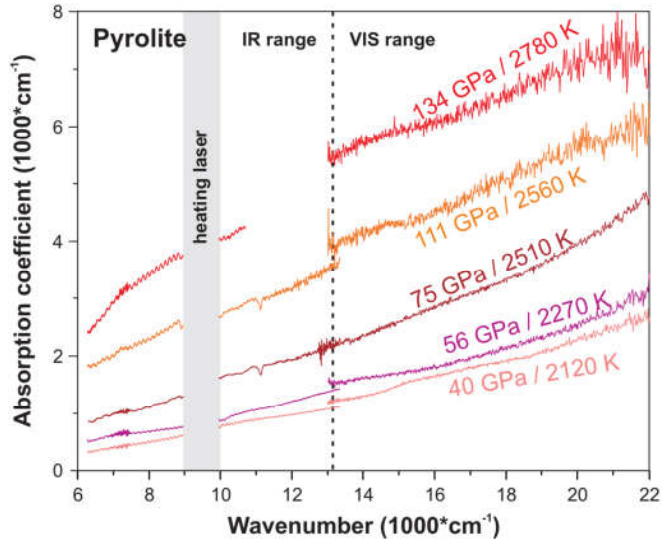

**Figure 2.** Pyrolite absorption coefficient at high *P-T* (before the scattering correction). Vertical dashed line separates the IR and VIS ranges as these were measured separately. The 9000-1000 cm$^{-1}$ region is blocked because of the need to filter out the 1070 nm heating laser radiation from reaching the detector. The 11000-13000 cm$^{-1}$ region in the 134 GPa / 2710 K spectrum (red) is not shown because of the poor signal-to-background ratio of the IR detector in this range at high absorbance levels.

The radiative conductivity of an absorbing medium is given by[6]: $k_{rad}(T) = \frac{4n^2}{3} \int_0^\infty \frac{1}{\alpha(v)} \frac{\partial I(v,T)}{\partial T} dv$ (Eq.2), where *α(v)* is the frequency-dependent absorption coefficient of the medium, *n* its refractive index, and *I(v, T)* is the Planck function. Prior to $k_{rad}$ evaluation, all absorption coefficients were corrected for light scattering in a submicron-grained sample based on the room-temperature absorption coefficients of single crystalline Bgm and Fp with corresponding compositions (Methods), which were measured separately using a different optical setup with a conventional (non-laser) optical probe[10] (Extended Data Fig. 7). Scattering-corrected absorption coefficients measured at *P-T* conditions approximating the geotherm yield $k_{rad}$ values that decrease continuously with depth from ~0.8 W/m/K at 1000 km to ~0.1 W/m/K at the top of the TBL (at ~2500 km) and then increases to ~0.35 W/m/K at the CMB (Fig. 3). Even smaller $k_{rad}$ values may be expected for the lower mantle if its grain size is smaller than the optical depth of pyrolite (< 100 μm) due to light scattering on grain boundaries. However, the grain size of the mantle is likely larger than the photon mean free path because of the hot regime favoring grain growth (*e.g.* Ref.[7]). The apparent negative slope of radiative thermal conductivity with depth is at odds with all previous radiative conductivity models, which show a positive trend and $k_{rad}$ up to 4-5 W/m/K at the CMB[7,8,11]. These previous models, however, were based on room-temperature measurements[7,9,11] or empirical considerations on the transparency of silicates at high temperature that assumed negligible temperature effect on the mantle optical properties[8]. Here we showed that pyrolite optical properties are highly sensitive to $T$ (Extended Data Fig. 4 and 5), suggesting that temperature-enhanced absorption coefficient leads to the negative $k_{rad}$ slope with depth.

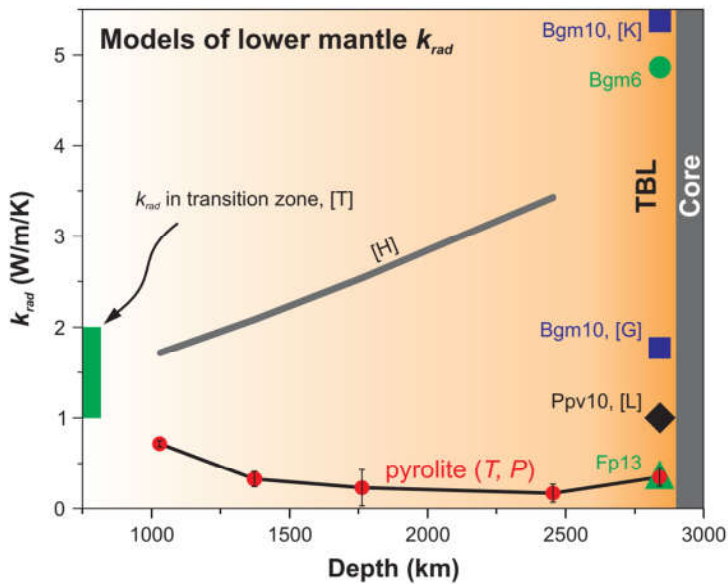

**Figure 3.** Radiative conductivity of the pyrolitic lower mantle along the geotherm[18] (red circles). The error bars vary with depth and reflect the quality of our scattering correction and uncertainty in the sample thickness at high pressure (Methods). The absorption coefficient for 2850 km was projected to 4000 K assuming a linear temperature-dependence of absorbance at $T > 2780$ K (Extended Data Fig. 6). Green circle and triangle are $k_{rad}$ values of Bgm and Fp with 6 and 13 mol.% Fe, respectively, based on their single crystal absorption coefficients measured at 117 GPa and 300 K (Extended Data Fig. 7). Grey thick curve marked [H] is the thermodynamic model of Ref.[8]: $k_{rad} = 1.9 * 10^{-10} * T^3$. Dark blue squares are $k_{rad}$ of Bgm with 10 mol.% Fe based on its 300 K absorption coefficient reported in Refs.[7,11], marked [K] and [G]. Black diamond [L] is $k_{rad}$ of Ppv with 10 mol.% Fe (Ref.[14]). Green bar [T] is the range of $k_{rad}$ values expected in the transition zone[13]. TBL is for thermal boundary layer above the core-mantle boundary. The color gradation illustrates the distribution of temperature in the lower mantle.

Light absorption in pyrolite at high *P-T* is governed by specific absorption mechanisms in its constituent phases. In order to disentangle the contribution of Bgm and Fp to the high-temperature spectra of pyrolite we measured the absorption coefficient of Bgm (single crystal, 6 mol.% Fe) up to ~2500 K (Extended Data Fig. 8). Surprisingly, the absorption coefficient of Bgm increases with the rate of ~0.1 cm$^{-1}$/K, while a much stronger temperature-dependence of

~0.4 cm$^{-1}$/K is characteristic of pyrolite (Extended Data Fig. 6). This observation hints that the strong increase in pyrolite absorbance at high temperature is not due to Bgm but is a result of the temperature-enhanced opacity of Fp. Unfortunately, we could not obtain reliable high-temperature absorption spectra of Fp as it showed irreversible changes in absorbance over continuous laser heating at $T > 1000$ K. We tentatively attribute this behavior to Soret-like iron diffusion under the steep temperature gradient typical of laser-heated samples. Cation diffusion rates in Fp are ~2 orders of magnitude faster than in Bgm[19], which may be why reversible spectroscopic behavior is preserved in Bgm upon continuous laser heating up to $T > 2500$ K. In the pyrolite sample iron diffusion is further limited by the small grain size and the reversible spectroscopic behavior is preserved up to ~3000 K (at 134 GPa).

To suppress iron diffusion and directly compare the temperature-induced changes in the Bgm and Fp optical absorbance we decreased the total laser-heating time by a factor of ~10$^6$ in an independent series of dynamically-heated DAC experiments. Fully reversible optical absorbance was observed in Bgm and Fp over the single-shot 1 μs long laser-heating to $T > 3000$ K at 111-135 GPa (Extended Data Fig. 9). Upon heating to ~3000 K at 135 GPa, Fp becomes ~5 times opaquer than it is at room temperature. In contrast, Bgm is only 20-30 % opaquer at ~3500 K than at room temperature. Because these transformations are fully reversible, they must reflect temperature-induced changes in light absorption by these single crystals. Our new streak camera data unambiguously show that light absorption in Fp is much more sensitive to temperature than in Bgm, confirming that the optical opacity of the lowermost mantle, and by extension its radiative conductivity, is governed by Fp. Such a diverse high-temperature behavior of Fp and Bgm likely results from the different mechanisms that govern light extinction in these phases. Even at room temperature single crystalline Fp with 13 mol.% Fe at 117 GPa is 2-3 times opaquer than Bgm with 6 mol.% Fe due to the intense Fe-O charge transfer band that extends into the VIS range. Lattice vibrations in hot Fp dynamically decrease the Fe-O separation, red-shifting the Fe-O charge transfer band, narrowing the band gap, and serving as a physical mechanism that effectively blocks radiative heat transport in the lowermost mantle. The proposed mechanism is much less effective in Bgm which accommodates iron predominantly at the larger (dodecahedral) site with longer average Fe-O distances than in Fp.

The high-pressure high-temperature behavior of Fp is of key importance for a pyrolytic model, indicating that Fp plays a major role in the opacity of the lower mantle and the effectiveness of radiative heat transport. In addition, the decreased band gap enhances electrical conductivity of Fp in the narrow depth range of the TBL (< 100 km) where the temperature increase is sharpest. Our interpretation of the optical behavior of Fp at high $P$-$T$ is in qualitative agreement with the computational evidence for a very high electrical conductivity (~4*10$^4$ S/m) in Fp at CMB conditions due to iron *d*-states forming broad bands near the Fermi level[20]. As such, our work hints that the physical properties of Fp contribute complexity to the CMB region and are responsible for some of its exotic features. For example, temperature-enhanced electrical conductivity of Fp in the TBL can account for the apparent absence of lag between the fluctuation in day length and geomagnetic jerks, which requires a thin (< 50 km) layer of highly conducting mantle just above the core[21].

Firmly constrained total thermal conductivity at the base of the mantle offers valuable insights into the CMB heat flow through the Fourier law of heat conduction: $Q_{CMB} = A_{CMB} * k_{total} * \Delta T$, where $A_{CMB}$ is the surface area of the CMB and $\Delta T$ is the temperature gradient above the CMB. Experimental estimates of the lattice thermal conductivity for a 4:1 mixture of Bgm (or Ppv) and Fp at CMB conditions fall in the range of 7.7-8.8 W/m/K (Extended Data Table 3)[2,3,22]. While these values are based on measurements in DACs at 300 K and rely on extrapolations to CMB temperatures, they are consistent with *ab initio* calculations of lower mantle thermal conductivity[23] as well as a recent experimental study of the lattice conductivity in Fe- and Fe,Al-bearing Bgm[24]. Recent estimates of the total thermal conductivity at the base of the mantle accepted $k_{rad}$ of ~2-4 W/m/K (*e.g.* Refs.[2,3]). The results of this work clearly indicate that radiative conductivity is not an important mechanism of heat transport in the lowermost mantle and may be safely neglected from estimates of the total thermal conductivity at the CMB. Accepting $k_{total} \approx k_{lat} \approx$ 8.5 W/m/K and a temperature gradient of ~0.007 K/m in the TBL[18], we obtain a $Q_{CMB}$ of ~ 8.5 TW for the present-day CMB.

This rate of heat extraction out of the core is sufficient to drive the present-day geodynamo, thanks to the compositional buoyancy due to the inner core growth, and is consistent with the inner core young age of ~ 0.6-1.3 Gy (*e.g.* Ref.[25]), assuming radioactive heat production is small in the core. A much older Earth's magnetic record (Hadean-Paleoarchean) requires a geodynamo prior to the inner core nucleation, operating in a convecting liquid core. Two distinct scenarios follow from the low mantle thermal conductivity. If core thermal conductivity is relatively low (< 50 W/m/K)[26,27], a CMB heat flux that is greater than the core adiabatic heat flux enables a thermally-driven geodynamo in the ancient fully molten core[5]. Alternatively, high core thermal conductivity (> 100 W/m/K)[4,28] is inconsistent with a thermally-driven geodynamo, and requires a compositionally-driven convection powered by light element exsolution upon core cooling[29-31]. Even if an early geodynamo can be driven in the cooling ancient molten core, determining its exact driving mechanism requires a definitive assessment of Earth's core thermal conductivity. Beyond Earth's magnetic field, our results shed light on the plausibility of long-lived dynamos in other terrestrial planets and exoplanets, and could provide a discriminating argument for their past or present habitability.

**Acknowledgements**

SSL acknowledges the support of the Helmholtz Young Investigators Group CLEAR (VH-NG-1325). This work was supported by the NSF Major Research Instrumentation program DMR-1039807, NSF EAR-1520648 and NSF EAR/IF-1128867, the Army Research Office (56122-CH-H), the National Natural Science Foundation of China (grant number 21473211), the Chinese Academy of Sciences (grant number YZ201524), the Carnegie Institution of Washington and Deep Carbon Observatory. The authors acknowledge T. Okuchi and N. Purevjav for their assist with the synthesis of the brigdmanite sample at Okayama University at Misasa. Parts of this work were conducted at GSECARS 13IDD beamline of the Advanced Photon Source, Argonne National Laboratory. GeoSoilEnviroCARS (Sector 13), Advanced Photon Source (APS), Argonne National Laboratory is supported by the National Science Foundation – Earth Sciences (EAR-1128799) and Department of Energy – GeoSciences (DE-FG02-94ER14466). This research used resources of the Advanced Photon Source, a U.S. Department of Energy (DOE) Office of Science User Facility operated for the DOE Office of Science by Argonne National Laboratory under Contract No. DE-AC02-06CH11357.

## Methods

### Pyrolite glass synthesis

Pyrolite glasses were synthesized using a gas-mixing aerodynamic levitation laser furnace at IPGP (Paris). Individual high-purity components ($CaCO_3$, $MgO$, $Al_2O_3$, $SiO_2$, $Fe_2O_3$,) of a CMASF-pyrolite composition were mixed together and ground with ethanol in an agate mortar. The powder was then dried and decarbonated in a furnace at 850 °C for 12 hours. This starting mix was pressed into pellets and broken into ~10 mg chunks, which were fused together in a gas-mixing aerodynamic levitation laser furnace at 2000 °C for 120 seconds in controlled $fO_2$ conditions and quenched to glass. The gas mixture used for levitation was 4.8% $CO_2$, 3.2% $H_2$ and Ar corresponding to a low $fO_2$ of IW+0.7 (0.7 log units above the iron-wustite buffer).

### Diamond anvil cell (DAC) assemblage

Three types of sample assemblages were used in this study. In the first type (40, 56, and 75 GPa runs), two holes were laser-drilled in Re gaskets pre-indented to 30-40 μm by diamond anvils with 200 and 300 μm culets. One of the holes was 50-70 μm in diameter and was filled with the pyrolite glass. The second hole was 30-40 μm in diameter and was filled with KCl serving as the optical reference for absorbance measurements at high pressure (see below). In the 111-134 GPa runs (second type of DAC assemblage), Re gaskets were indented to ~20 μm by anvils equipped with 80/300 μm beveled culets and only one hole (~40 μm in diameter) was laser-drilled in the center of the indentation serving as a sample chamber. In this setup the unheated glass was used as the optical reference because it is more transparent than heated pyrolite (Extended Data Fig. 3). Here, the resulting absorption spectra were corrected for the glass optical absorption determined in a separate run with a polished pyrolite glass sample loaded in the KCl pressure medium to 135 GPa (third type of DAC assemblage). In all DAC assemblages pressure was measured by the diamond Raman edge method with an uncertainty of ~5-10% (Ref.[32]).

### Optical measurements at high pressure and temperature

Optical absorption measurements in the IR (6200-13000 $cm^{-1}$) and VIS (13000-22000 $cm^{-1}$) ranges were performed separately using two laser-heating systems equipped with different spectrometers and detectors. The probe light source used in these measurements was the same: Leukos Pegasus pulsed supercontinuum laser (4 ns pulse at 0.25-1 MHz repetition rate) with a continuous spectrum in the 400-2400 nm (~4000-25000 $cm^{-1}$) range. The probe light was inserted into the optical paths of the IR and VIS systems, focused to ~5 μm in diameter, and aligned to the heating laser. The heating laser in both IR and VIS setups was a 1070 nm quasi-continuous Yt-doped fiber laser with improved beam quality[33]. Each spectral collection in IR or VIS was initiated by a 1 s heating pulse and offset from the start of the heating cycle by 200 ms.

In the IR runs we used a liquid nitrogen cooled ungated InGaAs detector (Princeton Instruments Model 7498-0001) coupled with the Acton Spectra Pro 2300i spectrometer with a 150 gr/mm grating. The probe collection window and the supercontinuum repetition rate were varied depending on the sample temperature in order to decrease the contribution of thermal background. At $T > 1500$ K, for example, typical collection windows were decreased to 1-5 ms and the probe frequency increased to 1 MHz. Despite these efforts, at $T > $~2700-3000 K the intensity of thermal radiation dominated in the collected signal, precluding accurate absorbance measurements with the ungated detector in the IR range. The immense brightness of the probe, however, allows acceptable signal-to-background in the IR range at $T < $~2700 K (Extended Data Fig. 10A). Spectral collections were performed at two spectrograph positions and then stitched for the combined spectrum in the 6200-13000 cm$^{-1}$ range.

In the VIS experiments, the transmitted probe light was collected by a 300 mm focal length spectrometer equipped with a 300 gr/mm grating and a gated (30 ns gate width at ~41.6 kHz repetition rate) iCCD detector (Andor iStar SR-303i-A). The detector gates were synchronized with the probe (fixed to 250 kHz for VIS measurements) such that the gate opens up to receive one supercontinuum pulse and then remains closed before the next sixth (due to the probe/gate frequency mismatch) probe pulse arrives. This synchronization scheme allows blocking the detector from most of the thermal radiation of the heated sample in between the supercontinuum pulses (see Ref.[34] for timing details). Similar to the IR setup, the collection window was decreased down to 1-10 ms for measurements at $T > $~2000 K to minimize the contribution of thermal background (Extended Data Fig. 10B). As is clear from Extended Data Fig. 10, the use of the gated detector allows eliminating thermal contribution in the VIS range at $T < $~2700 K. Spectral collections were performed at five spectrograph positions and then stitched for the combined spectrum in the 13000-22000 cm$^{-1}$ range.

Absorption coefficient in the VIS and IR at high $T$ was evaluated as $\alpha(v) = \ln(10) * \frac{1}{d} * (-log_{10}\left(\frac{I^T_{sample} - Bckg^T}{I^{300\,K}_{reference} - Bckg^{300\,K}}\right))$, where $d$ is the sample thickness, $I^T_{sample}$ is the intensity of light transmitted through the sample at $T$, $I^{300\,K}_{reference}$ is the intensity of light through the pressure medium at 300 K, $Bckg^T$ is background at $T$, and $Bckg^{300K}$ is background at 300 K. Background at $T$ was collected at the same laser-heating power and timing setup right after the

high-*T* absorption measurement but with the blocked probe light. In addition, the reported pyrolite absorption coefficients were corrected for absorbance of untransformed glass based on the absorption coefficients of crystallized pyrolite determined independently in experiments with thermal insulation between the sample and diamond anvils. We note, however, that this correction was small (<5 % of the total light extinction) because such a glass layer is thin (~1 µm)[35] and much more transparent than crystallized pyrolite (Extended Data Fig. 3). We neglected the reflections on the sample-medium interface as these are small (< 1 %) due to the similarity of the KCl and pyrolite refractive indices ($n$~2)[36]. The gap in absorption coefficients at ~9000-10000 cm$^{-1}$ is due to the IR filters in front of the IR spectrometer in order to block the 1070 nm heating pulse. Room-temperature pyrolite absorption coefficient measured with a non-supercontinuum (conventional) light probe (*e.g.* Ref.[10]) were used to linearly extrapolate the high-temperature absorption coefficients collected with the supercontinuum probe down to 3000 cm$^{-1}$.

Dynamic experiments were performed using the same laser heating system as in the VIS experiments but using a Sydor Ross 1000 streak camera coupled to a Princeton Instruments spectrometer (f/4, 150 grooves/mm) to record spectroscopic information. Single-pulse laser heating (1 µs) together with time-resolved probing with the supercontinuum laser (1 MHz, 1 ns long pulses) allowed collecting all spectroscopic information over one laser-heating cycle. Typical collection window was set to 30 µs with a heating laser arriving at the 8$^{th}$ µs. DACs cool back to 300 K in a matter of ~ 10 µs after the arrival of the heating pulse[37], thanks to the very high thermal conductivity of diamond anvils. Accordingly, we could collect up to 10 spectra that are characteristic of distinct thermal states of the samples (Extended Data Fig. 9). Details of this setup are provided in Ref.[38].

**Sample thickness**

Accurate absorption coefficients require that the sample thickness is known. To this end, we used two independent techniques yielding consistent thickness estimates. First, sample thickness was determined using visible light interferometry[39] at high pressure in the KCl cavity (first type of DAC assemblage). Refractive index of KCl was assumed to have a linear density dependence[36]. Here, overall thickness uncertainty is < 10-20 % and is mostly associated with the ambiguity in the KCl refractive index and uneven gasket thickness in the first DAC assemblage. The second estimate was obtained using the Zygo NewView 5032 optical 3D profilometer on recovered samples at 1 atm (Extended Data Fig. 11). This technique allows high precision (~10 nm) imaging of the surface roughness and, in principle, must yield a more accurate estimate than the first technique because of the fewer assumptions involved. The thickness at high pressure was then obtained by correcting the Zygo-based estimate using the bridgmanite $P$-$V$-$T$ equation of state[40]. A simple elastic behavior was assumed for the sample on decompression, which is a reasonable approach because plastic deformations on decompression are typically small[39], following previous high-pressure studies where sample thickness is a critical parameter (*e.g.* Refs.[2,41]). The difference between the two estimates was within 10-20 %. Nonetheless, in evaluating absorption coefficients we used Zygo-based estimates as they have less intrinsic uncertainties and are more precise.

**Double-sided laser-heating and temperature measurements**

Two separate laser-heating systems were used for IR and VIS runs with laser-heating spots being ~4 (IR) and ~2 times (VIS) wider than the probe spot, which ensures that the probed area is uniformly heated. Both IR and VIS setups allow for double-sided laser-heating. However, only the VIS setup allows double-sided temperature measurements. In the IR runs, thermal emission was collected on the downstream side (with respect to the supercontinuum) by the same InGaAs detector as described above with the grating position centered at 700 nm. For the VIS, we followed the previously described procedure for temperature evaluation[34]. Optical responses of the IR and VIS systems were calibrated using a standard white lamp (Optronics Lab OL 220C) and the collected blackbody radiation was fitted using the T-Rax software (C. Prescher) to extract the temperature value. The reported temperatures, such as shown in Figure 1, are the average of IR and VIS measurements that were within 50 K (Extended Data Fig. 10). Our overall temperature uncertainty is typical of laser-heated diamond anvil cells and is on the order of 150-200 K (Refs.[42,43]).

**Light scattering correction**

The observed submicron grain size (Fig. 1) suggests that light losses due to scattering may be significant. If such, the measured absorption coefficient is: $\alpha(v)^{measured} = \alpha(v)^{intrinsic} + \alpha(v)^{scattering}$, where $\alpha(v)^{intrinsic}$ is the intrinsic absorption coefficient due to light-induced excitation processes in the sample. The magnitude of light scattering can be estimated as:
$\alpha(v)^{scattering} = \alpha(v)^{measured} - x * \alpha(v)^{bridgmanite} - (1 - x) * \alpha(v)^{ferropericlase}$, where $\alpha(v)^{bridgmanite/ferropericlase}$ is the 300 K absorption coefficient of single-crystal bridgmanite or ferropericlase with chemical composition typical of that in pyrolite, and $x = 0.8$ is the fraction of bridgmanite in the pyrolitic lower mantle[1]. Here we assume that light scattering in single crystals is small and that light absorption in the lower mantle is limited to bridgmanite and ferropericlase, which is reasonable as the iron content of Ca-perovskite is negligible. Absorption coefficients of doubly polished single crystal bridgmanite (6 mol.% Fe) and ferropericlase (13 mol.% Fe with 10% $Fe^{3+}/Fe_{total}$) with compositions approximating that observed by TEM in pyrolite (Extended Data Table 2) were measured separately using our conventional IR-VIS light absorption system (*e.g.* Ref.[44]) up to 130 GPa at room temperature. Please note that the $Fe^{3+}$ content of our ferropericlase single crystals determined via electron energy loss spectroscopy (10% $Fe^{3+}/Fe_{total}$) is in the range of that reported for ferropericlase crystals synthesized at high pressure in equilibrium with bridgmanite (5-19 $Fe^{3+}/Fe_{total}$)[45,46] and also close to that measured directly in our pyrolite sample after its crystallization at ~56 GPa (~5-17 $Fe^{3+}/Fe_{total}$).

Extended Data Fig. 7 compares absorption coefficient of doubly polished single crystalline bridgmanite and ferropericlase to that of pyrolite (measured with the same optical setup). If static light scattering on grain boundaries in the polycrystalline sample was negligible, then its absorption coefficients should be close to that of a mixture of 80 % bridgmanite with 20 % ferropericlase. From Extended Data Fig. 7 it is clear that light scattering on grain boundaries is significant and increases with frequency. We, thus, estimate the frequency-dependent static light scattering as described above and use the obtained scattering coefficients to correct the spectra shown in Fig. 2 for light scattering in the polycrystalline pyrolite samples.

In applying the scattering correction we assumed that static light scattering on grain boundaries is the dominant scattering mechanism. At high temperature, however, dynamic light scattering can also come into play near or above melting temperatures, which is why it has been used as a melting criterion in some high-pressure studies[47]. Please note that all spectroscopic measurements were performed at temperatures that are at least 1000 K degrees lower than the expected melting temperatures of pyrolite. Furthermore, we observe a near-linear increase in temperature-induced absorbance up to ~3000 K (Extended Data Fig. 6), which is consistent with thermally-activated absorption but inconsistent with dynamic light scattering. Finally, temperature-enhanced absorbance is fully reversible, which clearly would not be the case should grains change their orientation and/or size considerably upon heating. For these reasons, we can safely rule out dynamic light scattering as an alternative explanation for the observed temperature-enhanced light absorption.

**Light scattering in Fe-free pyrolite**

Independently, we synthesized Fe-free pyrolite glass (similarly to the above procedure) and subjected it to $CO_2$-laser-heating to > 2000 K at 48 GPa. The crystallinity of the sample was confirmed by synchrotron XRD. Interestingly, we were also able to detect bridgmanite in the Raman spectra of Fe-free pyrolite, which indirectly confirms our explanation for the absence of

Raman signal in Fe-bearing pyrolite (Extended Data Fig. 1) as being due to strong light absorption in the Fe-bearing sample.

Crystallized samples of Fe-free pyrolite allowed us to measure light scattering directly in the DAC at high pressure because no light is absorbed within the sample by the *d-d* transitions or charge transfer in the IR-VIS range. Static light scattering coefficients obtained for Fe-free pyrolite at 48 GPa are 10-50 % (IR-VIS) lower than our estimate based on the single crystals at 40 GPa. One plausible reason to this deviation is that the single crystal scattering model is based on bridgmanite and ferropericlase samples that are slightly iron depleted compared to that in pyrolite. Nevertheless, this result serves as an independent test of the quality of the scattering correction used in this study.

**Computational support for light scattering correction**

The scattering of light is modeled with the superposition T-matrix method using the program Multiple Sphere T-matrix, which is an exact solver of Maxwell equations for a system of spherical particles[48]. The scattering medium is 10 x 10 x 5 μm rectangular volume filled with scattering particles (Extended Data Fig. 12). Each particle is assumed to be a sphere with a diameter of 0.4 μm. The packing density of the particles in this medium is 0.2. The medium and particles are given real parts of the index of refraction of 2.15 and 2.00, respectively, approximating the expected values at 135 GPa (Extended Data Fig. 13). Three different choices for the imaginary part of refractive index were investigated: 0.000, 0.031, and 0.062 where 0.031 and 0.062 represent a common value for iron oxide minerals in the visible to near-IR range at ambient conditions[49,50]. A beam of light with a wavelength of 400 and 800 nm is shined to the medium from the negative to positive z direction at (x,y) = (0,0).

Our computations show that light scattering in a finely-grained diamond anvil cell sample is predominantly in the forward direction with scattering angles of 0-20 degrees, where the scattering angle is the angle between the incident light vector and the scattered light vector (Extended Data Fig. 12). Taking into account the numerical aperture of the objective used to collect the transmitted portion of the radiation (NA = 0.4) and diamond refractive index (2.417), we obtain that ~60 % of the total scattered light is sampled by the detector. Notably, our results for 400 and 800 nm incident light differ only by 3-5 %, depending on the choice of refractive index. These estimates indicate that the amount of light collected by the detector is determined by absorption and scattering events in comparable extents.

Importantly, our computations provide an upper bound on the scattered light analyzed by the detector because of light losses due to reflections on grain boundaries and due to the more complex (non-spherical) real morphology of the mineral interfaces. On the other hand, our scattering correction based on the absorption coefficients of Bgm and Fp single crystals yields that ~15-25 % of scattered light reaches the detector. This provides a lower bound on the scattering because the iron content of the samples used to estimate light absorption (on *d-d* and charge-transfer transitions) was ~20 % lower than that in the crystallized pyrolite (Extended Data Table 2). As such our empirical scattering correction likely overestimates the actual scattering, implying that pyrolite radiative conductivity at mantle *P-T* may even be smaller than the presented estimate by ~50 %. This results in the $k_{rad}$ uncertainty due to the ambiguity in light scattering of ~0.07 W/m/K at the core-mantle boundary, well within the overall uncertainty in $k_{rad}$ of ~0.5 W/m/K estimated by us and 10-100 times smaller than the current ambiguity in the total thermal conductivity of the lower mantle.

**FIB sample preparation and TEM measurements on pyrolite after optical measurements at high *P-T***

A thin section suitable for TEM analysis was extracted from the sample equilibrated at 56 GPa and ~ 2800 K using the FIB lift-out technique, using a Zeiss Auriga 40 instrument at IPGP (Paris). First, a 1.5-µm-thick platinum layer was deposited on top of the sample, to prevent ion beam damage of the sample during excavation. The sample was excavated around the platinum deposit using a focused gallium ion (Ga+) beam operated at 30 kV using milling currents between 20 nA and 1 nA. The thin section was subsequently extracted with the help of a micromanipulator and transferred onto an Omniprobe TEM copper grid for further thinning down to electron transparency. For the latter, we used lower milling currents (80 pA to 700 pA) than for excavating the sample. Since TEM analyses require a sample thickness below 100 nm, and given the uncertainty in thickness measurement while ion imaging, we thinned the sample until we reached electron transparency in SEM imaging with the electron beam operated at 4 kV; this proved to be sufficient enough to provide good quantitative analyses for TEM-EDX. The final step in the sample preparation consists in polishing the thinned area to remove any deposit from the milling process. For this, we operated the focused ion beam at 5 kV and used a current of 30 pA.

Imaging and chemical analysis of the sample were performed with a FEI Tecnai Osiris transmission electron microscope equipped with four windowless Super-X silicon drift detectors (SSD), provided by Bruker, for energy-dispersive x-ray (EDX) spectroscopy. The microscope was operated at 200 kV with 1.2 nA probe current in scanning transmission electron microscopy

(STEM) mode. EDX maps were acquired over 1024x1024 pixel areas for 800-1000 seconds per map using a 50 microseconds pixel dwell time. EDX spectrums from the selected grains were processed and quantified with Cliff-Lorimer method integrated in the Esprit 1.9 software package from Bruker.

### Electron Energy Loss Spectroscopy (EELS)

In order to compare the $Fe^{3+}$ content of ferropericlase in pyrolite samples to that used in ferropericlase single crystals used for scattering correction, we performed EELS analysis on the FIB-cut foils (< 100 nm thick). The foils were prepared by a Helios G4 UC DualBeam-system from FEI at GFZ. The FIB lamella had a dimension of 15 x 6 x 0.1 µm. The rough mill occurred with an acceleration voltage of 30 kV and a beam current of 47 nA to 9.3 nA. The rough cut out samples were lifted out with an Easyliftout system and fixed onto a half-moon copper-grid. There these were thinned to a thickness of 0.1 µm with a beam current from 2.5 nA to 80 pA. The final polishing was done with 41 pA and 5kV.

EELS data were acquired using a Tecnai F20 X-Twin TEM operated with a field emission gun electron source. Acceleration voltage was 200keV. Energy-loss spectroscopy was performed utilizing a Gatan Tridiem™ system. The conditions for spectra acquisition in TEM diffraction mode were: camera length = 700 mm; 0.1 eV/channel; collection aperture = 2mm; convergence angle = 4 mrad; collection angle 8 mrad; 20 spectra with 1 second acquisition time each were added up to one spectrum; energy resolution = 1.09 eV; deconvolution of the spectra using zero loss peak (zlp). All spectra were processed and deconvoluted using the Gatan Digital Micrograph software package. At least, 10 spectra from different locations were acquired from one sample. The EELS data were analyzed following the procedure outlined by van Aken and Liebscher [51] with the reported $Fe^{3+}$ fraction ($Fe^{3+}/Fe_{total}$) being an average across the different locations. The uncertainty is typical of EELS measurements of $Fe^{3+}$ content.

### Uncertainties in evaluated pyrolite radiative conductivity

Two main sources contribute uncertainty to our radiative conductivity model: (i) ambiguity associated with the scattering correction and (ii) an unknown pyrolite mean refractive index. Below we describe our approach to estimating the total uncertainty in our $k_{rad}$ values.

The total ambiguity in scattering coefficient is difficult to assess because it results from the mismatch in chemical composition (iron content, its oxidation, and crystallographic position) of single crystalline bridgmanite and ferropericlase (used to estimate the absorption coefficient of pyrolite with no scattering on grain boundaries) with that in our pyrolite samples (finely grained). In addition, single crystalline samples are always imperfect[52]; thus, cannot be viewed as strictly non-scattering media. Whether the concentration of scattering centers in the single crystals studied here is comparable to that in the lower mantle is unknown. Nevertheless, the contribution of intrinsic defects in single crystals to the total light scattering in a finely grained polycrystalline sample is likely negligible[17]. We have estimated the quality of our scattering correction by comparing $k_{rad}$ values for the finely grained pyrolite (scattering-corrected) to that of the composite single crystalline model using 300 K absorption coefficients (Extended Data Fig. 7). We find that the mantle $k_{rad}$ models based on these two sets of absorption coefficients are within 0.1-0.4 W/m/K. Accordingly, we reflect this in our estimate of the total uncertainty in $k_{rad}$ (Fig. 3).

Refractive indices of bridgmanite and ferropericlase have never been measured as a function of pressure or temperature. In order to evaluate their mean values we employed the approach of Ref.[53], which allows calculating refractive indices of a mineral based on the empirically established polarizabilities of its constituent ions (*e.g.* $Mg^{2+}$, $Fe^{2+}$, $Si^{4+}$, $O^{2-}$) and its molar volume (*i.e.* density). Assuming simplified chemical compositions of Bgm, $(Mg_{0.92}Fe_{0.08})SiO_3$, and Fp, $(Mg_{0.82}Fe_{0.18})O$, for the phases analyzed by TEM in this work (Extended Data Table 2) and using corresponding 300 K equations of state for $MgSiO_3$ (Ref.[40]) and MgO (Ref.[54]) we obtain mean refractive indices of these phases as a function of pressure at room temperature (Extended Data Fig. 13). Based on this model, the uncertainty in the obtained pyrolite refractive index is less than 5 % (Ref.[53]). Accordingly, we employed a pressure-dependent mean pyrolite refractive index for our computations of $k_{rad}$. Please note that possible discontinuities in pyrolite density over spin transitions in Bgm and Fp were not taken into account in our density estimation for the following reasons. First, spin transition in Bgm, which accounts for ~0.8 fraction of the lower mantle, has not been identified in samples with compositions representative of the lower mantle. In fact, Bgm with realistic Al and Fe does not undergo spin transition[55]. Second, the spin transition in Fp with the composition representative of the pyrolitic model (~18 mol.% Fe) takes place at 40-60 GPa at room temperature[56]. However, the density variation associated with the spin transition is rather small (2-3 %)[57,58]; thus, would not contribute significantly to the refractive index. Considering, the fraction of Fp in the lower mantle (~0.2) and assuming a change in density of 2-3 %, the associated change in the bulk refractive index of pyrolite would be ~0.5 %, which we do take into account in our overall uncertainty in refractive index of ~5 %.

Temperature-induced variations in *n* are not important. For example, in diamond variation of *n* is ~ 1%/1000 K (Ref.[59]). Such information is completely lacking for Bgm and Fp at relevant *P-T* conditions. As such, we used Gladstone-Dale relation and established thermal equations of state of MgO and $MgSiO_3$ (Refs.[60,61]) to estimate the change in *n* upon heating to 3000 K. For both MgO and $MgSiO_3$ the corresponding change in *n* is ~1 % and of the same sign (both have positive thermal expansion). The overall uncertainty in pyrolite *n* is within 10 %, which contributes an uncertainty of < 20 % to the $k_{rad}$ value.

### Comparison to previous estimates of radiative thermal conductivity

Previous room-temperature high-pressure studies of polycrystalline Bgm with 10 mol.% Fe have reported values of ~1.5 (Refs.[9,11]) and ~5.5 W/m/K (Ref.[7]) at near CMB conditions. In comparison, the absorption coefficient of our single crystalline Bgm with 6 mol.% Fe collected at 300 K and 117 GPa (Extended Data Fig. 7) yields $k_{rad}$ of ~5 W/m/K at CMB conditions (Fig. 3). Please note that here we used the absorption coefficient collected at 300 K to allow the comparison with previous studies. Our $k_{rad}$ value for single crystalline Bgm with lower Fe content plots in between the values reported for polycrystalline Bgm with higher Fe content[7,11], suggesting that the previous $k_{rad}$ estimates[7,9] suffer from both inaccurate thickness measurements and light scattering in the polycrystalline samples. Unlike previous studies, in this work we measured the sample thickness by two independent techniques, which yielded consistent results, as well as used single crystalline Bgm. In consistency with the previous optical studies of Bgm[7,9,11] we confirm that its absorption coefficient shows relatively insignificant variations with pressure. Unlike Bgm, Fp absorption coefficient is highly sensitive to pressure and shows an apparent discontinuity over the spin transition (Extended Data Fig. 7) in agreement with previous reports on optical properties of Fp at high pressure[15,16]. These examples show that the room-temperature absorption coefficients of Bgm and Fp measured in this work are consistent with the previous reports, which underscores the importance of temperature-induced variations in the mantle optical properties revealed in this work for mantle radiative thermal conductivity.

**Extended Data Figures**

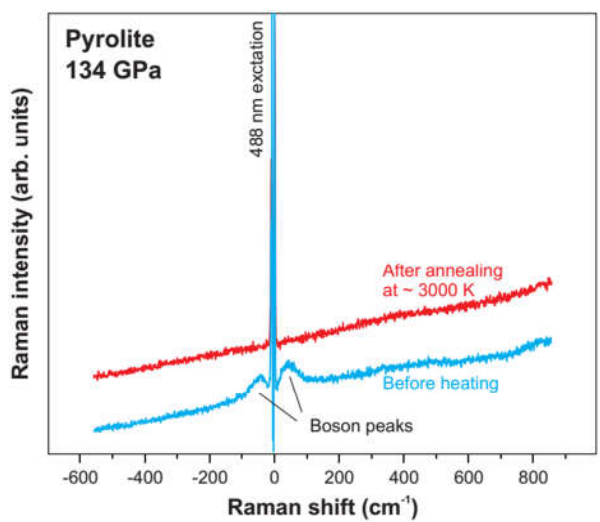

**Extended Data Figure 1.** Raman spectra of pyrolite at 134 GPa taken before (blue) and after crystallization at ~3000 K. Boson peaks, which are characteristic of glasses[62], are not observed in the heated sample suggesting its full crystallinity. Please note that while the corresponding x-ray diffraction clearly shows the presence of bridgmanite, post-perovskite, and ferropericlase after the annealing (Extended Data Fig. 2), we could not resolve their Raman features. Importantly, we can expect only bridgmanite and post-perovskite to give Raman signatures as ferropericlase does not have Raman active modes. We propose that ferropericlase, which is present as small inclusions, causes a substantial reduction of the laser probe and Raman scattering intensity due to its large light extinction coefficient (as is shown in this work). The absence of Raman features of brdgmanite after pyrolite crystallization has previously been observed and attributed to the small grain size and/or higher symmetry due to the incorporation of Al- and Fe-, which would lower its Raman activity[63]. However, we did observed Raman bands of $MgSiO_3$-perovksite after the crystallization of Fe-free pyrolite (see below for details), which confirms indirectly that the absence of bridgmanite in the Raman spectrum after the crystallization of Fe-bearing pyrolite is due to exceptional ferropericlase opacity.

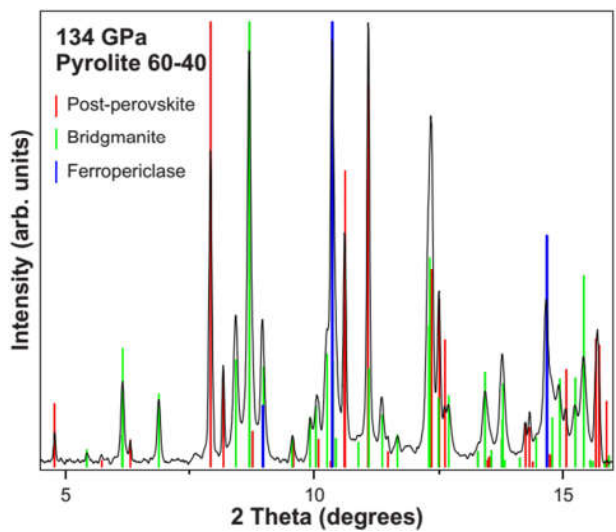

**Extended Data Figure 2.** X-ray diffraction pattern of a pyrolite glass after laser-heating to ~ 3000 K at 134 GPa. X-ray wavelength is 0.3344 Å.

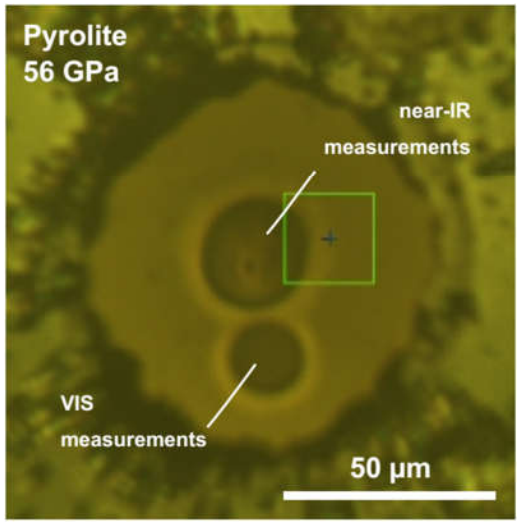

**Extended Data Figure 3.** Optical image of the pyrolite glass laser-heated in two sample spots for near-IR and VIS optical measurements at 56 GPa and up to ~ 2800 K. Unheated pyrolite glass appears more transparent.

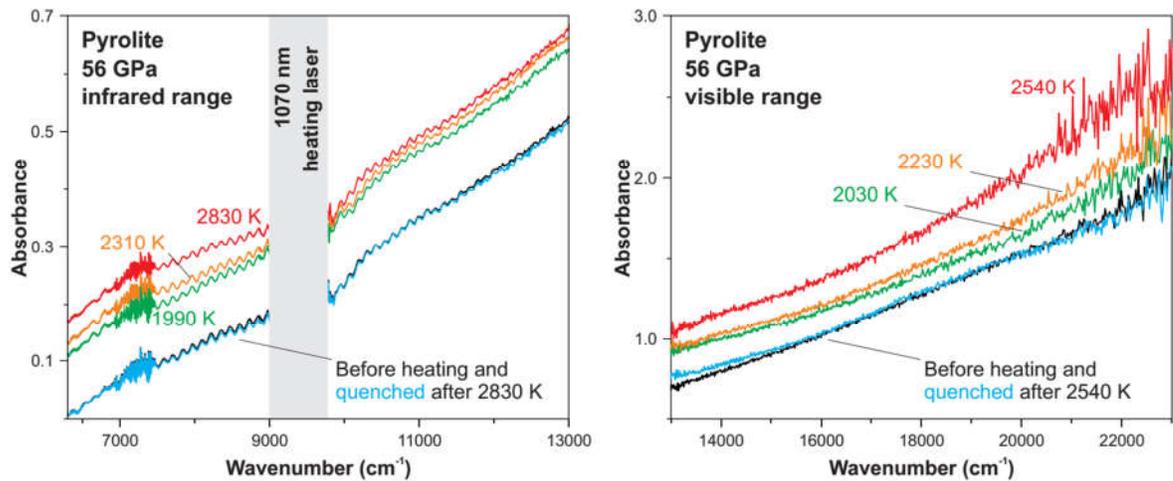

**Extended Data Figure 4.** Pyrolite absorption spectra (before corrections) at 56 GPa and high $T$ as recorded by the IR and VIS systems. The spectra were evaluated as $Abs(v) = -log_{10}\left(\frac{I^T_{sample} - Bckg^T}{I^{300\,K}_{reference} - Bckg^{300\,K}}\right)$, where $I^T_{sample}$ is the intensity of light transmitted through the sample at $T$, $I^{300\,K}_{reference}$ is the intensity of light through the reference at 300 K, $Bckg^T$ is background at $T$, and $Bckg^{300K}$ is background at 300 K. Please note that the before and after heating spectra are almost identical, indicating that the $T$-induced changes in sample absorbance are completely reversible. Low absorbencies at < ~7000 cm$^{-1}$ are due to the small size of the hole used to collect the optical reference. All spectra used to compute radiative conductivity were corrected for that by referencing to high-quality spectra collected on a double-polished pyrolite sample crystallized at similar $P$-$T$ conditions.

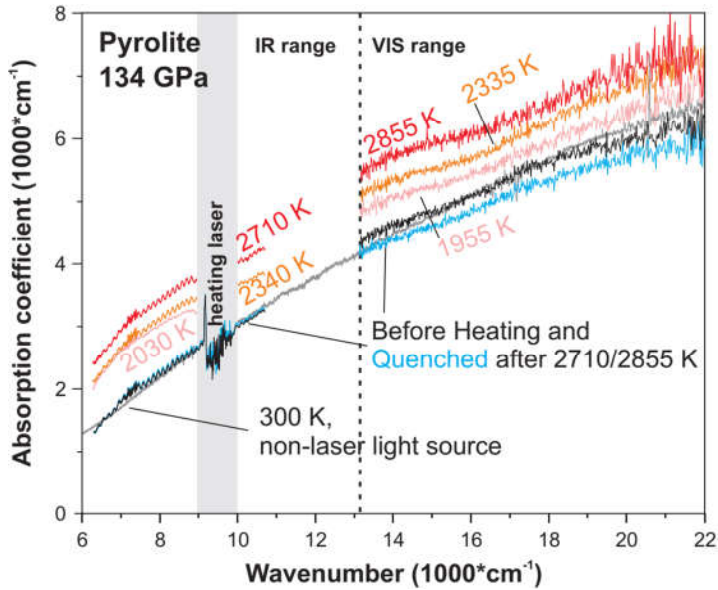

**Extended Data Figure 5.** Absorption coefficients of pyrolite at 134 GPa) and high *T* as recorded by the IR and VIS systems (before scattering correction). The spectra were evaluated as

$\alpha(v) = -\ln(10) * \frac{1}{d} \log_{10}\left(\frac{I^T_{sample} - Bckg^T}{I^{300\,K}_{reference} - Bckg^{300\,K}}\right)$, where $I^T_{sample}$ is the intensity of light transmitted through the sample at $T$, $I^{300\,K}_{reference}$ is the intensity of light through the reference at 300 K, $Bckg^T$ is background at $T$, $Bckg^{300K}$ is background at 300 K, and $d$ is sample thickness. Black and blue are spectra measured before heating and after quenching from 2710 and 2855 K in the IR and VIS runs, respectively. The grey spectrum was collected using the conventional optical absorption system (no supercontinuum used) (*e.g.* Ref.[44]) and is shown for comparison.

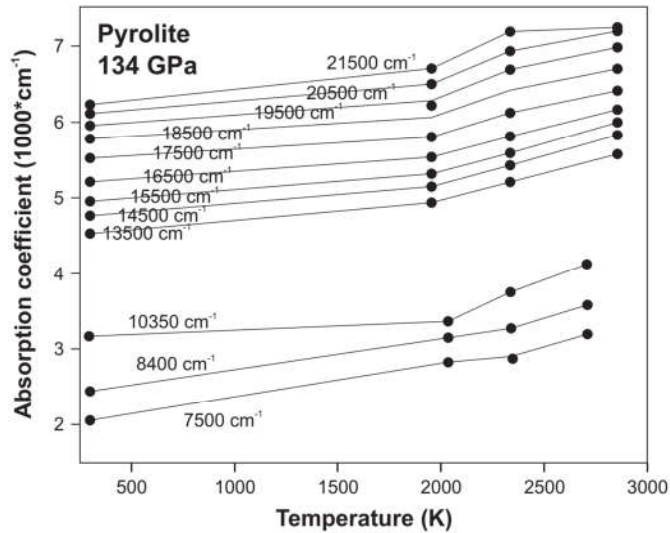

**Extended Data Figure 6.** Temperature-dependence of pyrolite absorption coefficients (before scattering correction) at selected frequencies and 134 GPa (black circles). The absorption coefficient increases with temperature as ~0.35-0.45 cm$^{-1}$/K (depending on frequency).

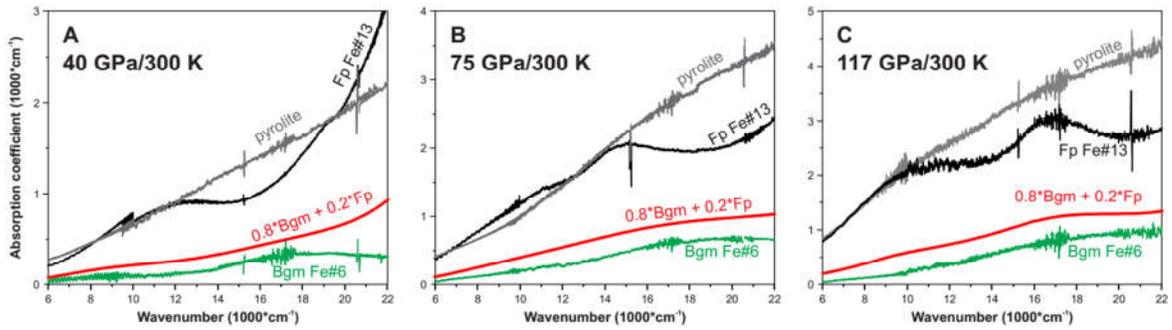

**Extended Data Figure 7.** Room-temperature absorption coefficients of pyrolite (grey) as compared to that of doubly polished bridgmanite (green) and ferropericlase (black) single crystals at 40 (**A**), 75 (**B**), and 117 GPa (**C**). The composition of Bgm and Fp was chosen to approximate that in pyrolite as revealed by our TEM measurements. Red spectra are hypothetical absorption coefficients of a rock with 80% Bgm and 20% Fp. The frequency-dependent light scattering was estimated by subtracting the red spectra from grey. All spectra were measured using the same optical setup with an identical light source. Likewise, thicknesses of all these phases were measured using the Zygo profilometer at 1 atm and optically (at high pressure), as described in the text.

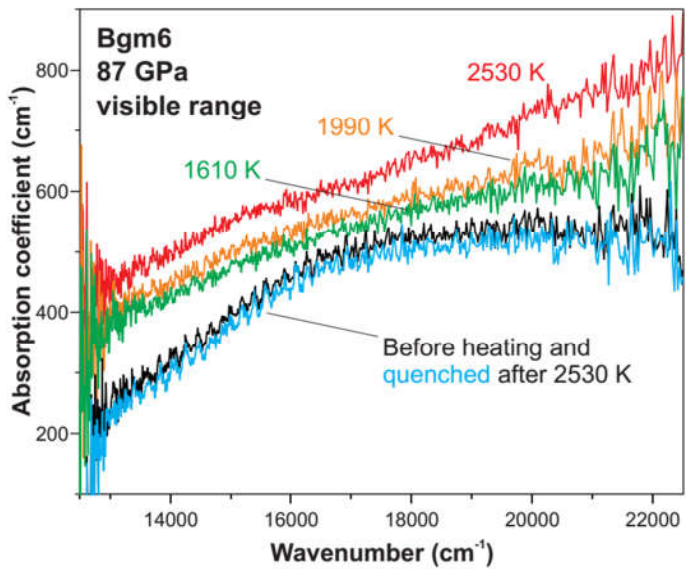

**Extended Data Figure 8.** Absorption coefficient of single crystal bridgmanite with 6 mol.% Fe (same as in Extended Data Fig. 7) at 87 GPa at room temperature (black and blue), 1610 K (green), 1990 K (orange), and 2530 K (red). Absorption coefficient increases with temperature as ~0.07-0.15 cm$^{-1}$/K (depending on frequency).

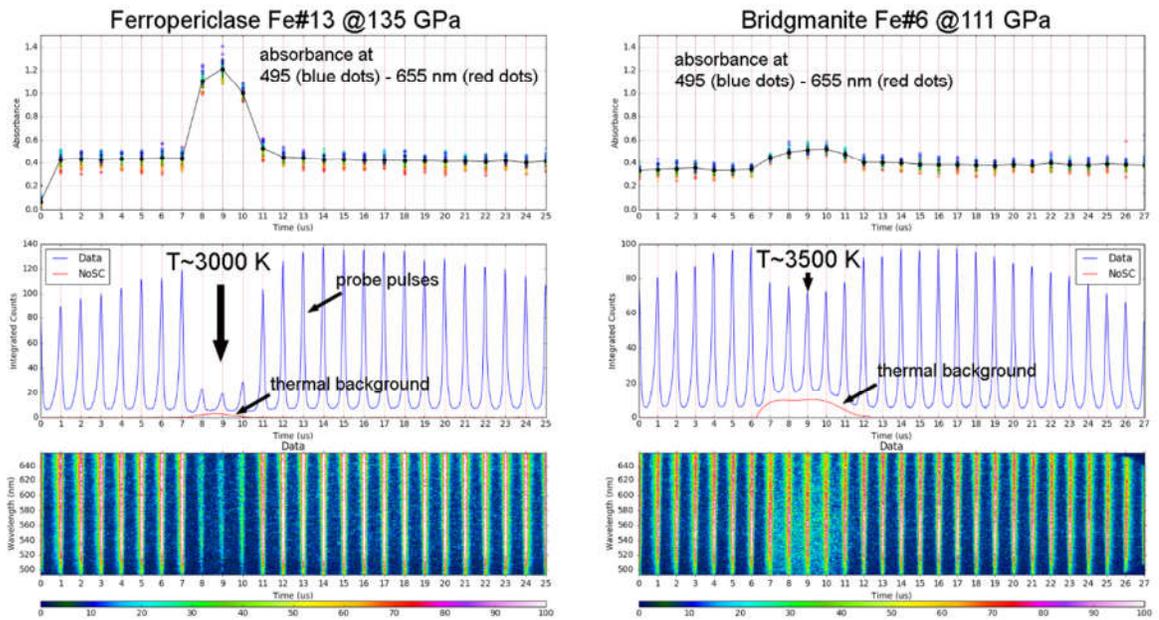

**Extended Data Figure 9. Top panels:** Evolution of optical absorbance of ferropericlase (Fe#13) and bridgmanite (Fe#6) at 135 and 111 GPa upon one-microsecond laser heating as measured by a streaked detector, similar to our recent observation of metallic nitrogen[64]. The 1μs heating laser arrives between the 7-th and 8-th microsecond. Colored dots average absorbance in the 495 (blue dots) – 655 (red dots) nm range (each dot corresponds to a range of ~10 nm). Please, note the reversibility of the spectra over the heating. **Middle panels:** Detector counts integrated over all wavelengths in the 495-655 nm range. Note the striking difference between the high-temperature absorbance in Fp and in Bgm. **Lower panel:** Raw streak camera data. Vertical lines are the supercontinuum pulses traversing the sample every microsecond.

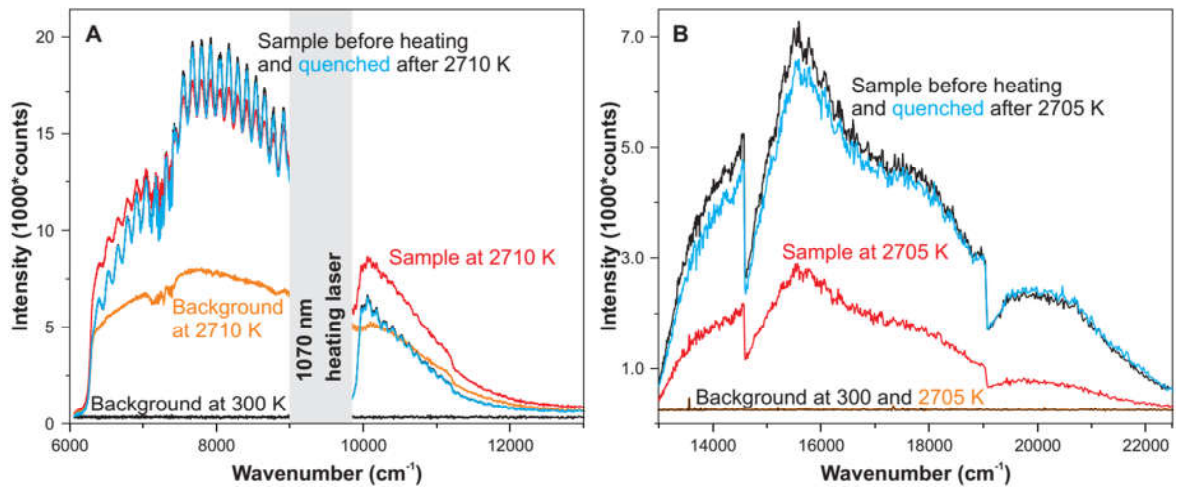

**Extended Data Figure 10.** Intensity of the supercontinuum light transmitted through the crystallized pyrolite at 300 K (black and blue) and 2710/2705 K (red) at 134 GPa in the IR (**A**) and VIS (**B**) ranges. Please note the difference between the thermal background accumulated for ~1000 probe pulses traversing the sample at ~2700 K in the IR (ungated detector) versus VIS (gated iCCD).

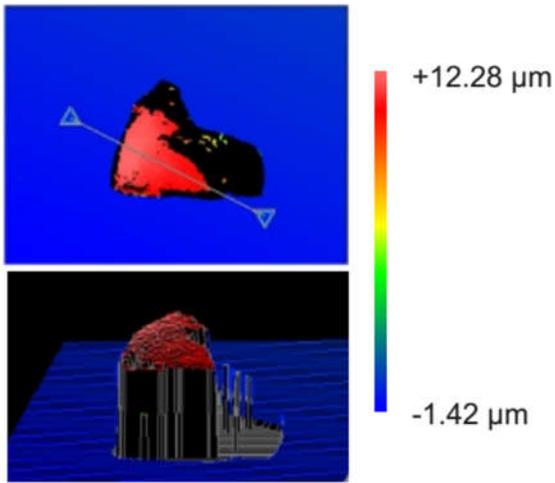

**Extended Data Figure 11.** An example of the thickness measurement using Zygo NewView 5032 3D optical profilometer on a recovered pyrolite sample at 1 atm after its decompression from 40 GPa.

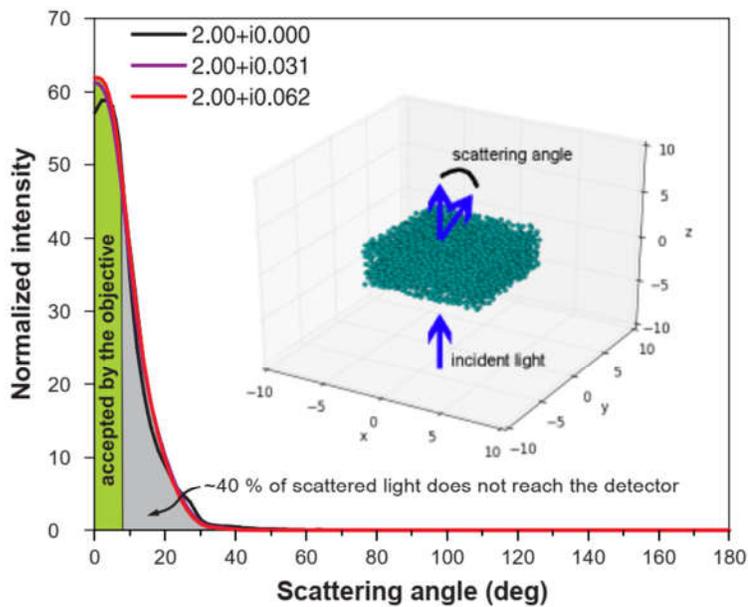

**Extended Data Figure 12.** Scattering angles as modeled for the sample geometry shown in the inset, which approximates sample geometry at $P > 100$ GPa. Three different curves depict the effect of variable imaginary part of refractive index. **Inset:** Light scattering was computed for a box of 10x10x5 μm filled with 400 nm (DI) spherical scattering particles with a packing density of 0.2, representing ferropericlase, and a medium, representing brdigmanite.

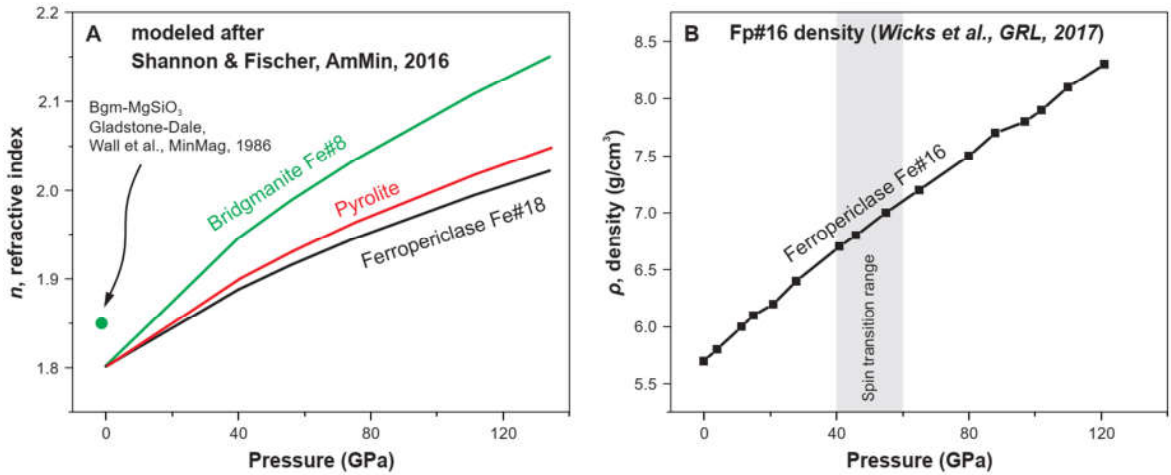

**Extended Data Figure 13.** (**A**) Room-temperature mean refractive indices of bridgmanite-$(Mg_{0.92}Fe_{0.08})SiO_3$ and ferropericlase-$(Mg_{0.82}Fe_{0.18})O$ as a function of pressure evaluated following Ref.[53]. Green circle is the refractive index of iron-free bridgmanite computed in Ref.[65] using the Gladstone-Dale relation, shown for comparison. Pyrolite refractive index was evaluated as: $n^{pyrolite} = 0.8*n^{bridgmanite} + 0.2*n^{ferropericlase}$. We estimate the uncertainty in pyrolite refractive index at 300 K as less than 5 % (Ref.[53]). (**B**) Density of ferropericlase with 16 mol.% Fe (Ref.[58]) shows no apparent discontinuity over the spin transition range at 40-60 GPa.

### Extended Data Tables

**Extended Data Table 1.** Chemical compositions of the pyrolite glass in wt.%.

|  | C | O | MgO | SiO$_2$ | CaO | FeO | Al$_2$O$_3$ |
|---|---|---|---|---|---|---|---|
| **Mean value** | 0 | 0 | 38.15 | 46.60 | 2.16 | 8.66 | 4.24 |
| **Sigma** | 0 | 0 | 0.12 | 0.10 | 0.03 | 0.12 | 0.06 |

**Extended Data Table 2.** Chemical compositions of coexisting ferropericlase and bridgmanite grains crystallized from a pyrolite glass laser-heated at 56 GPa up to ~ 2800 K.

| Ferroperclase | Mg | Fe | Al | Si | Ca | Fe# |
|---|---|---|---|---|---|---|
| map1 | 82.34 | 16.05 | 0.36 | 0.97 | 0.28 | 16.31 |
|  | 81.52 | 16.96 | 0.72 | 0.71 | 0.09 | 17.22 |
| map2 | 81.52 | 15.92 | 1.69 | 0.81 | 0.06 | 16.34 |
|  | 81.77 | 15.51 | 1.51 | 1 | 0.21 | 15.94 |
|  | 81.58 | 15.52 | 1.29 | 1.34 | 0.26 | 15.98 |
| map3 | 80.89 | 18.07 | 0.09 | 0.94 | 0.01 | 18.26 |
|  | 81.42 | 17.22 | 0.61 | 0.76 | 0 | 17.46 |
|  | 80.69 | 17.08 | 0.35 | 1.85 | 0.02 | 17.47 |
|  | 80.18 | 17.93 | 0.28 | 1.52 | 0.09 | 18.28 |

| | | | | | | | |
|---|---|---|---|---|---|---|---|
| map4 | | 80.15 | 18.77 | 0.03 | 1.04 | 0 | 18.97 |
| | | 80.88 | 18.06 | 0.07 | 1 | 0 | 18.25 |
| map5 | | 83.26 | 14.43 | 1 | 1.25 | 0.06 | 14.77 |
| | | 82.82 | 14.99 | 1.06 | 1.12 | 0 | 15.33 |
| | | 82.31 | 15.65 | 0.81 | 1.14 | 0.1 | 15.98 |
| | | 82.73 | 15.48 | 0.69 | 1.09 | 0.01 | 15.76 |
| Average | | **81.60** | **16.51** | **0.70** | **1.10** | **0.08** | **16.83** |
| | | | | | | | |
| **Bridgmanite** | | Mg | Fe | Al | Si | Ca | Fe# |
| map1 | | 40.86 | 3.02 | 5.84 | 47.96 | 2.32 | 6.88 |
| | | 41.91 | 3.27 | 5.39 | 47.25 | 2.18 | 7.24 |
| | | 41.43 | 2.89 | 5.43 | 47.92 | 2.33 | 6.52 |
| map2 | | 40.79 | 3.17 | 4.97 | 48.7 | 2.38 | 7.21 |
| | | 41.1 | 3.31 | 4.99 | 48.21 | 2.38 | 7.45 |
| | | 39.84 | 3.05 | 6.13 | 48.06 | 2.92 | 7.11 |
| map3 | | 40.75 | 3.59 | 5.13 | 48.54 | 1.99 | 8.10 |
| | | 40.85 | 3.52 | 5.22 | 48.51 | 1.9 | 7.93 |
| | | 39.51 | 3.94 | 4.88 | 48.82 | 2.85 | 9.07 |
| | | 40.74 | 3.5 | 4.83 | 48.55 | 2.38 | 7.91 |
| map4 | | 39.63 | 4.21 | 5 | 48.38 | 2.79 | 9.60 |
| | | 39.45 | 4.2 | 5.56 | 48.37 | 2.43 | 9.62 |
| map5 | | 40.84 | 3.26 | 5.2 | 48.43 | 2.27 | 7.39 |
| | | 41.42 | 2.82 | 4.95 | 48.66 | 2.14 | 6.37 |
| | | 40.69 | 3.06 | 5.46 | 48.55 | 2.24 | 6.99 |
| | | 40.73 | 3.44 | 5.01 | 48.32 | 2.5 | 7.79 |
| Average | | **40.66** | **3.39** | **5.25** | **48.33** | **2.38** | **7.70** |
| | | | | | | | |
| **Ca- perovskite** | | Mg | Fe | Al | Si | Ca | |
| map5 | | 3.37 | 1.01 | 1.03 | 48.05 | 46.54 | |

**Extended Data Table 3.** Aggregate lattice thermal conductivity of pyrolite at the core-mantle boundary *P-T* conditions deduced from conductivities of individual minerals using the Hashin-Shtrikman bounds.

| Phase | k, W/m/K | Rock | Reference | Method |
|---|---|---|---|---|
| Fe-free pyrolite, lat | 11.0 ± 2.0 | Bgm-pyrolite | Ohta, et al. [41] | Thermoreflectance |
| Fe-free pyrolite, lat | 17.8 ± 3.9 | Ppv-pyrolite | Ohta, et al. [41] | Thermoreflectance |
| Pyrolite, lat | 7.7 ± ? | Bgm-pyrolite | Dalton, et al. [22] | Thermoreflectance |
| Pyrolite, lat | 7.9 ± 1.3 | Bgm-pyrolite | Ohta, et al. [2] | Thermoreflectance |
| Pyrolite, lat | 8.8 ± 1.1 | Bgm-pyrolite | Okuda, et al. [3] | Thermoreflectance |
| | | | | |
| Fe-free pyrolite, lat | 4.0 ± 0.5 | Bgm-pyrolite | de Koker [66] | *ab initio* |
| Fe-free pyrolite, lat | 16.5 ± ? | Bgm/Ppv-pyrolite | Haigis, et al. [67] | *ab initio* |
| Fe-free pyrolite, lat | ~2 ± ? | Bgm-pyrolite | Tang, et al. [68] | *ab initio* |
| Pyrolite, lat | 8.1 ± 1.1 | Bgm-pyrolite | Stackhouse, et al. [23] | *ab initio* |

## Extended References